\documentclass[twocolumn,showpacs,amsmath,amssymb,prl]{revtex4}
\usepackage{graphicx}
\usepackage{dcolumn}
\usepackage{bm}

\begin{document}

\title{Magnetoelectric Effects on Composite Nano Granular Fe/TiO$_{2-\delta}$ Films}

\author{S.D. Yoon}
\altaffiliation[Corresponding author e-mail: s.yoon@neu.edu ]{}
\author{C. Vittoria}
\author{V.G. Harris}
\affiliation{Center for Microwave Magnetic Materials and Integrated Circuits, 
Department of Electrical and Computer Engineering, Northeastern University, 
Boston, MA. 02115 USA}

\author{A. Widom}
\affiliation{Department of Physics, Northeastern University, Boston, MA. 02115 USA}

\author{Y.N. Srivastava}
\affiliation{Physics Department \& INFN, University of Perugia, Perugia, IT}

\begin{abstract}
Employing a new experimental technique to measure magnetoelectric 
response functions, we have measured the magnetoelectric effect 
in composite films of nano granular metallic iron in anatase titanium 
dioxide at temperatures below 50 K. A magnetoelectric resistance is 
defined as the ratio of a transverse voltage to bias current 
as a function of the magnetic field. In contrast to the anomalous Hall 
resistance measured above 50 K, the magnetoelectic 
resistance below 50 K is significantly larger and  
exhibits an even symmetry with respect to magnetic field reversal $H\to -H$. 
The measurement technique required attached electrodes in the plane of the 
film composite in order to measure voltage as a function of bias current 
and external magnetic field. To our knowledge, the composite films are unique 
in terms of showing magnetoelectric effects at low temperatures, $<$ 50 K, 
and anomalous Hall effects at high temperatures, $>$ 50 K.
\end{abstract}


\maketitle
Present research in new functional materials, such as magnetoeletric (ME) 
or multiferroic (MF) materials, have drawn a great deal of attention because of 
their potential applications in electronics and nano electronic device technologies\cite{Fiebig:2005,Kimura:2003,Wang:2003,Hur:2004,Lottermoser:2004,Spaldinr:2005,Eerenstein:2006}. Properties of ME and MF materials have been recently reviewed\cite{Fiebig:2005,Lottermoser:2004,Spaldinr:2005,Eerenstein:2006}.
The effects are a consequence of the coupling between electric and magnetic fields 
in materials. ME and MF materials have multi purposes or multi functional applications. 
Relatively few multiferroic materials exist in natural phases, such as TbMnO$_3$,\cite{Fiebig:2005,Kimura:2003} BiFeO$_3$,\cite{Fiebig:2005,Wang:2003,Zavaliche:2005} HoMnO$_3$,\cite{Hur:2004}. 
Composite materials combining dielectric and 
ferromagnetic materials have been suggested as possible ME or MF materials\cite{Fiebig:2005,Lottermoser:2004,Spaldinr:2005,Eerenstein:2006}. 
Synthesis of composite materials having ME or MF properties are now of great laboratory interest\cite{Fiebig:2005,Kimura:2003,Wang:2003,Hur:2004,Eerenstein:2006,Huang:2007}.

In our previous work\cite{Yoon:2006,Yoon1:2007}, oxygen deficient anatase structure 
titanium dioxide TiO$_{2-\delta}$ films on (100) lanthanum aluminates LaAlO$_3$  
were deposited. The films exhibited both ferromagnetic and semiconducting properties 
at room temperature. In order to enhance the saturation magnetization at room temperature 
in these semiconducting films, we have incorporated  nano granular (NG) metallic iron (Fe) spheres in highly epitaxial oxygen deficient anatase TiO$_{2-\delta}$.
The resulting composite films gave rise to some intriguing phenomenon that 
the composite films exhibited strong carrier spin polarization 
of anomalous Hall effects (AHE) at $T >$ 200 K, where the carrier density 
was measured to be $n > 10^{22} {\rm /cm^3}$. 

\begin{figure}[tp]
\centering
\includegraphics[width=0.48\textwidth]{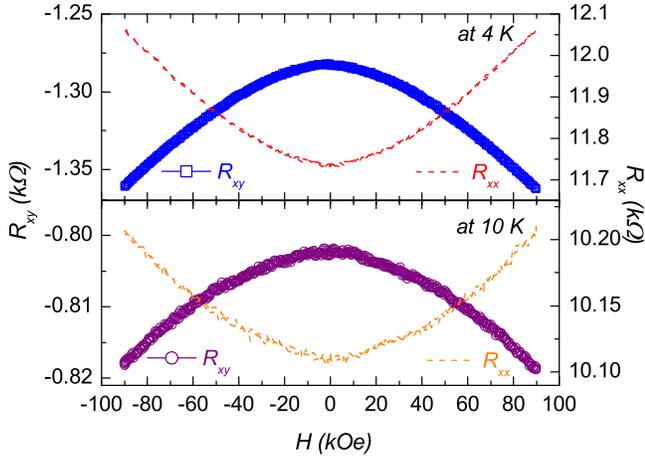}
\caption{Magnetoresistance $R_{xx}(\bf H)$ (dashed lines) and Magnetoelectric resistance 
$R_{xy}(\bf H)$ (solid symbols) 
measured in function of magnetic field at two temperatures $T <$ 50 K.}
\label{Resistance} 
\end{figure}

We now report magnetoelectric effects in the films for $T <$ 50 K. These were measured 
by a conventional four probe Hall transport measurement technique. The transverse voltage, $V_\perp$, was measured perpendicular to the bias current $I$. 
We define the ME resistance, $R_{xy}$, as $R_{xy} =(V_\perp)/I$. In FIG. \ref{Resistance}, 
we plot the measured $R_{xy}(\bf H)$ as a function of $H$. The bias current $I$ was fixed 
as the magnetic field was varied from $- 90\ {\rm kOe} < H < + 90\ {\rm kOe}$. The bias 
current was applied in the film plane. FIG. \ref{Resistance} shows that the measured 
$R_{xy}({\bf H})$ is an even function of ${\bf H}$, where ${\bf H}$ was applied normal 
to the film plane. It is argued below that the measured voltage is due to the ME effect 
of the film rather than the Hall  effect, since $R_{xy}({\bf H})$ is an even 
function of ${\bf H}$. Indeed, when we  apply ${\bf H}$ in the film plane, where there 
is no possibility of Hall voltage to be  measured in the film plane, $R_{xy}(\bf H)$ nevertheless 
behaves similarly to FIG. \ref{Resistance}. 
Thus, for temperatures below 50 K, the composite films are 
characterized by a magnetoelectric effect and above 50 K by an 
anomalous Hall effect\cite{Yoon2:2007}.

A standard technique\cite{Baques:2006,Devan:2006} for measuring the ME effect in films is 
to monitor the electric polarization via voltage measurements across the film thickness 
in the presence of an external magnetic field ${\bf H}$. This technique may be indirect 
if the ME film material is, for example, deposited on a nonmagnetic substrate. 
We have devised a simple direct scheme by which the ME effect is measured in film materials 
by placing electrodes in the plane of the film rather than across the film. We utilize 
the same measurement technique as devised in Hall measurements whereby a bias current is applied in the film plane bisecting the two electrodes. Whereas in the measurement of the Hall voltage ${\bf H}$ is applied normal to the film planes, in the measurements of ME voltage ${\bf H}$ may be applied in any direction. From the measured $I-V$ characteristics 
at fixed ${\bf H}$, one may deduce the magnetization as a function of electric fields which  is indeed the ME effect. There is a basic difference between the Hall measurement and the 
ME measurement technique as devised here, and they are: (i) Experimentally, the Hall  voltage can be measured only for ${\bf H}$ applied normal to the film plane, whereas in 
our present technique ${\bf H}$ may be applied in any direction. For example, for 
${\bf H}$  applied in the film plane, the electrodes in the film plane would not detect 
any Hall voltage. (ii) Theoretically, for a fixed current $I$, the Hall voltage is an odd  function of ${\bf H}$ and the ME voltage is an even function of ${\bf H}$. 

In previous work\cite{Yoon2:2007} the magnetoresistance $R_{xx}(T,{\bf H})$ 
and the Hall resistance $R_{xy}({\bf H})$ of the composite films were measured 
for temperatures  $4\ {\rm K} < T < 300\ {\rm K}$. For  $T > 50\ {\rm K}$, the anomalous Hall resistance ($R_{xy}({\bf H})$) and magnetoresistance $R_{xx}({\bf H})$, respectively, were measured. $R_{xy}({\bf H})$ 
was measured to be an odd function of ${\bf H}$ at temperatures above 50 K. However, for temperature below 50 K and for ${\bf H}$ in the plane as well as normal to the film 
plane of $R_{xy}({\bf H})$ dependence on H was parabolic as shown in 
FIG.(\ref{Resistance}). Plots of $R_{xx}({\bf H})$ show positive MR in that range of temperatures (below 50 K). We suggest that this behavior of 
$R_{xy}({\bf H})$ at low temperatures may be due to ME effects in the composite films. 
$R_{xy}({\bf H})$ in FIG. (\ref{Resistance}) demonstrates both broken time reversal and broken parity conservation symmetry as in a ME system. Note that the product symmetry (parity multiplied by time reversal)  
is unbroken. Previous theoretical studies\cite{Widom:1985,Widom:1982,Widom:1984} of 
the ME effect are recalled to explain the behavior of $R_{xy}({\bf H})$. In order to 
derive a relationship between $R_{xy}({\bf H})$  and the ME effect, let us first 
consider free energy per unit volume $F$. The thermodynamic laws read\cite{Fiebig:2005,Eerenstein:2006}
\begin{eqnarray}
F({\bf E},{\bf H},T)=F(-{\bf E},-{\bf H},T),
\nonumber \\ 
dF=-SdT-{\bf P}\cdot d{\bf E}-\mu_{0}{\bf M}\cdot d{\bf H},
\label{Free-energy}
\end{eqnarray} 
wherein ${\bf P}$ is the polarization, ${\bf E}$ the electric field, 
${\bf M}$ the magnetization, and ${\bf H}$ the magnetic field vector. 
Note the symmetry condition 
\begin{eqnarray}
{\cal A}_{ij}({\bf E},{\bf H},T)=-\frac{\partial^2 F({\bf E},{\bf H},T)}
{\partial E_i \partial H_j}={\cal A}_{ij}(-{\bf E},-{\bf H},T),
\nonumber \\ 
{\cal A}_{ij}({\bf E},{\bf H},T)=\mu_0 
\left(\frac{\partial M_j}{\partial E_i }\right)_{{\bf H},T}
=\left(\frac{\partial P_i}{\partial H_j }\right)_{{\bf E},T}, 
\nonumber \\ 
\alpha_{ij}({\bf H},T)=c\lim_{{\bf E}\to 0}{\cal A}_{ij}({\bf E},{\bf H},T)
=\alpha_{ij}(-{\bf H},T), 
\label{Free-Energy-Symmetry}
\end{eqnarray} 
wherein $c$ is light velocity and $\alpha_{ij}$ is the magnetoelectric tensor. 
If a small electric field $\delta {\bf E}$ is present in the plane of the film, 
then the magnetization change due to the electric field 
obeys\cite{Widom:1985,Widom:1984} 
\begin{equation}
R_{vac} \delta {\bf M}=\alpha \delta {\bf E},
\label{ME-Relation Equation}
\end{equation}
where $\alpha $ is the ME psuedoscalar coupling coefficient and $R_{vac}=c\mu_0$ 
is the vacuum impedance. One may associate a uniform magnetization ${\bf M}$ 
with a surface current per unit length ${\bf K}$ via 
\begin{equation}
{\bf K}_{\rm magnetic}={\bf M} \times {\bf n},
\label{Surface-Current Equation}
\end{equation}
wherein ${\bf n}$ is a unit vector normal to the surface.  
Eqs.(\ref{ME-Relation Equation}) and (\ref{Surface-Current Equation}) on 
the film surface yield a surface magnetoelectric conductance $g$ according to 
\begin{equation}
\delta {\bf K}_{ME\ {\rm effect}}=g\delta {\bf E} \times {\bf n}, 
\ \ \ \ {\rm where}\ \ \ \ g=\frac{\alpha }{R_{vac}}\ .
\label{ME-Conductivity Equation}
\end{equation}
In the experimental four probe configuration on the film surface, 
\begin{equation}
\begin{pmatrix}
\delta E_x \\ \delta E_y
\end{pmatrix}=
\begin{pmatrix}
R_{xx} & R_{xy} \\ 
R_{yx} & R_{yy}  
\end{pmatrix} 
\begin{pmatrix}
\delta K_x \\ \delta K_y
\end{pmatrix} 
\label{matrix-r}
\end{equation}
where $R_{yx}=-R_{xy}$, 
yields the central result for the experimental method; i.e.  
\begin{equation}
\alpha = R_{vac}g = \frac{-R_{vac}R_{xy}}{{R_{xx}}^2+{R_{xy}}^2}.
\label{ME-coupling Equation}
\end{equation}

\begin{figure}[tp]
\centering
\includegraphics[width=0.43\textwidth]{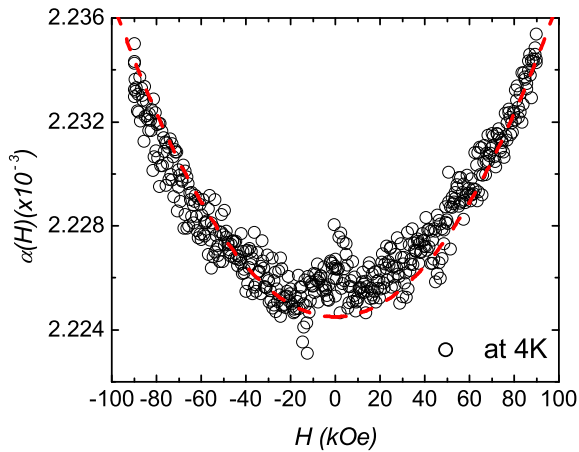}
\caption{Magnetroelectric coefficient, $\alpha(\bf H)$ (symbols), data 
from magnetoelectric conductivity. Dashed line shows fitting experimental 
ME coefficient with equation shown in Eq.(\ref{Fitting Equation}).}
\label{ME-coupling} 
\end{figure}

FIG. \ref{ME-coupling} exhibits typical $\alpha ({\bf H})$ data at $T = 4\ {\rm K}$ obtained from 
$R_{xx}$ and $R_{xy}$  versus ${\bf H}$. The theoretical fitting plot of $\alpha ({\bf H})$ 
was obtained according to 
\begin{equation}
\alpha({\bf H}) \approx 
\alpha(0)+\eta H^2.
\label{Fitting Equation}
\end{equation}
The fit between theory and experiment appears satisfactory.

In summary, magnetoelectric resistances of nano granular Fe in TiO$_{2-\delta}$ composite 
were measured as a function of magnetic fields and temperatures below 50 K. The ME coupling 
was shown to be related to the measured resistance $R_{xy}({\bf H})$ and $R_{xx}({\bf H})$ for temperatures 
below 50 K. Thermodynamic arguments were presented to explain the ME measurement technique, 
see FIG. (\ref{ME-coupling}). In view of the fact that in our composite films of nano granular 
metallic Fe spheres embedded in anatase TiO$_{2-\delta}$ we have a magnetostrictive material 
(nano granular Fe spheres) and piezoelectric materials (TiO$_{2-\delta}$), the resultant 
composite mimics the effects of ME materials. Our result appears  
consistent with results observed in other composite materials where magneto 
and ferroelectric material components were 
combined\cite{Fiebig:2005,Spaldinr:2005,Eerenstein:2006}. 
The magneto transport properties of our composite seems to suggest that there 
may be potential for spintronics and/or multifunctional nano electronic applications.

{\it This research was supported by the National Science Foundation (DMR 0400676) and 
the Office of Naval Research (N00014-07-1-0701).}

\end{document}